\documentclass{aastex61}
\usepackage{multirow}
\usepackage{captcont}
\received{November 6th 2018}
\revised{December 6th 2018}
\shorttitle{Rings across ages}
\shortauthors{van der Marel et al.}
\begin{document}

\title{Protoplanetary disk rings and gaps across ages and luminosities}

\correspondingauthor{Nienke van der Marel}
\email{astro@nienkevandermarel.com}

\author{Nienke van der Marel}
\affil{Herzberg Astronomy \& Astrophysics Programs, 
National Research Council of Canada, 
5071 West Saanich Road, 
Victoria BC V9E 2E7,
Canada}
\nocollaboration

\author{Ruobing Dong}
\affil{Physics \& Astronomy Department, 
University of Victoria, 
3800 Finnerty Road, 
Victoria, BC, V8P 5C2, 
Canada}
\nocollaboration

\author{James di Francesco}
\affil{Herzberg Astronomy \& Astrophysics Programs, 
National Research Council of Canada, 
5071 West Saanich Road, 
Victoria BC V9E 2E7,
Canada}
\nocollaboration

\author{Jonathan P. Williams}
\affil{Institute for Astronomy, 
University of Hawaii, 
2680 Woodlawn dr.,
96822 Honolulu HI,
USA}
\nocollaboration

\author{John Tobin}
\affil{Homer L. Dodge Department of Physics and Astronomy, 
University of Oklahoma, 
440 W. Brooks Street, 
Norman, OK 73019, 
USA}
\affil{Leiden Observatory, 
Leiden University, 
Niels Bohrweg 1, 
2333 CA Leiden, 
the Netherlands}
\affil{NRAO, 
520 Edgemont Road
Charlottesville, VA 22903-2475
USA}
\nocollaboration

\begin{abstract}
Since the discovery of the multi-ring structure of the HL~Tau disk, ALMA data suggest that the dust continuum emission of many, if not all, protoplanetary disks consists of rings and gaps, no matter their spectral type or age. The origin of these gaps so far remains unclear. We present a sample study of 16 disks with multiple ring-like structures in the continuum, using published ALMA archival data, to compare their morphologies and gap locations in a systematic way. The 16 targets range from early to late type stars, from $<$0.5 Myr to $>$10 Myr, from $\sim$0.2 to 40 $L_{\odot}$ and include both full and transitional disks with cleared inner dust cavities. Stellar ages are revised using new \emph{Gaia} distances. Gap locations are derived using a simple radial fit to the intensity profiles. Using a radiative transfer model, the temperature profiles are computed. The gap radii generally do not correspond to the orbital radii of snow lines of the most common molecules. A snow line model can likely be discarded as a common origin of multi-ring systems. In addition, there are no systematic trends in the gap locations that could be related to resonances of planets. Finally, the outer radius of the disks decreases for the oldest disks in the sample, indicating that if multi-ring disks evolve in a similar way, outer dust rings either dissipate with the gas or grow into planetesimal belts. 
\end{abstract}


\section{Introduction}
The HL~Tau image released as part of the first ALMA Long Baseline Campaign has mesmerized the astronomical and non-astronomical community \citep{HLTau2015}. The sharp rings and gaps seen in the dust continuum at 30 mas resolution have revolutionized the view of protoplanetary disks, which were previously assumed to be smooth at the early stages of evolution. The detailed structure in HL~Tau has generated a firestorm of theoretical work trying to explain the presence and multitude of gaps in a disk that is barely 1 Myr old, ranging from planets \citep{Dong2015gaps}, snow lines \citep{Zhang2015}, sintering \citep{Okuzumi2016}, instabilities \citep{Takahashi2016}, resonances \citep{Boley2017} to dust pile-ups \citep{Gonzalez2017}. Since HL~Tau, many other disks have been observed by ALMA at high spatial resolution, all revealing rings and gaps \citep[e.g.][]{Andrews2016,Isella2016,Fedele2017,Loomis2017,diPierro2018}. Interestingly, ring-like structures are found across the full ranges of spectral type, luminosity, and age, ranging from 0.4 Myr up to 10 Myr.   

The origin of the gaps and rings has so far remained unclear, especially considering the diversity in the morphologies. Gas surface density structures are the manifestation of dynamics in the disk and should in principle be able to distinguish between the different proposed mechanisms. CO observations have been used to derive the gas surface density structure in the gaps of HL~Tau and HD~163296 \citep{Yen2016,Isella2016}, and even CO kinematics have been derived inside the gaps of the latter \citep{Teague2018,Pinte2018}, suggesting the presence of planets. A depletion of dust grains, however, will change the gas temperature inside dust gaps \citep{Facchini2017gaps,vanderMarel2018-hd16}. Hence, care must be taken when interpreting CO data inside dust gaps.

Disk-planet interactions remain a common explanation for the observed ring-like structures. The discovery of such structures around stars that are less than 1 Myr old require an early formation of planets at the locations of the rings. Whether gas giant planets at tens of AU can form on such short timescale is unclear \citep{Helled2014}. Instead, snow lines \citep{Zhang2015} could explain the presence of gaps by enhanced dust growth beyond millimeter sizes due to increased dust grain stickiness at specific radii in disks. Alternatively, dust grains could grow to millimeter sizes at the snow line radii due to higher fragmentation velocities, and the gaps become depleted of millimeter grains due to radial drift down to the nearest snow line \citep{Cleeves2016, Pinilla2017-ice}. snow lines may even induce pressure bumps due to a change in ionization \citep{Flock2015}. A scenario involving snow lines would imply a clear relation between the stellar properties and the locations of the gaps, which has so far remained unconfirmed. 

If gaps are present in disks over a large range of ages, it raises the questions: when are they formed and what does that tell us about their origins? Also, how do the structures in young disks compare to those in older disks? Furthermore, several transition disks, disks with inner cleared dust cavities, have been shown to consist of multiple rings \citep[e.g.][]{Walsh2014,vanderMarel2015-12co,Fedele2017,vanderPlas2017}, suggesting these disks undergo similar processes as 'full disks' (defined as disks without inner cleared cavities) with rings. It remains to be understood how transition disks fit in the evolutionary picture of disk evolution. \citet{vanderMarel2018} proposed an evolutionary scenario, where disk rings as dust traps are the start of planetary growth, forming planets in the inner part of the disk that eventually clear out a large inner cavity. 

So far, most disk ring studies have focused on individual objects, using images at different spatial resolution and distance, and different model descriptions to fit the intensity profiles. In order to understand the origin of the rings, a large sample and a uniform analysis are required. In this study we present the analysis of 16 published ALMA continuum images of disks showing multiple gaps. Both full disks and transition disks are included, as the outer rings in some transition disks appear to follow a similar structure as their full counterparts. The chosen targets cover a range of stellar properties, environments (isolated/star forming regions) and ages, using new distance information from the recent \emph{Gaia} Data Release DR2 \citep{Gaia2018}. We aim here to compare the ring and gap locations with the stellar radiation field, disk age and disk mass to understand their common origin.

\section{Sample}
Our sample consists of 16 dust continuum disks observed and published so far that show two or more spatially resolved rings at millimeter wavelengths (either Band 6 or 7, 1.3 mm or 870 $\mu$m) known to date. We also include the disks HD~135344B and V1247 Ori, which both show an inner ring and an outer asymmetry, as the asymmetry can be considered to be part of a ring as well, triggered by the Rossby-Wave Instability. The majority of these images have been previously published and analyzed in detail (see Table \ref{tbl:sample}). When both bands are available at similar spatial resolution, we have chosen to use the Band 6 image, as the emission is more optically thin. Only in the case of RXJ1615.3-3255 we have selected the Band 9 data of program 2011.0.00724.S as their spatial resolution is higher than that of the Band 7 program 2012.1.00870.S. All disks are known to be actively accreting and have relatively high disk masses of $>$20 $M_{\rm Jup}$. The sample includes both full and transition disks with inner cavities. The disks reported as transition disks are AA~Tau, RXJ1615.3-3255, DM~Tau, V1247~Ori, HD~97048, HD~100546, TW~Hya, HD~135344B and HD169142. 

\subsection{ALMA observations}
\begin{figure}[!ht]
\centering
\includegraphics[width=\textwidth]{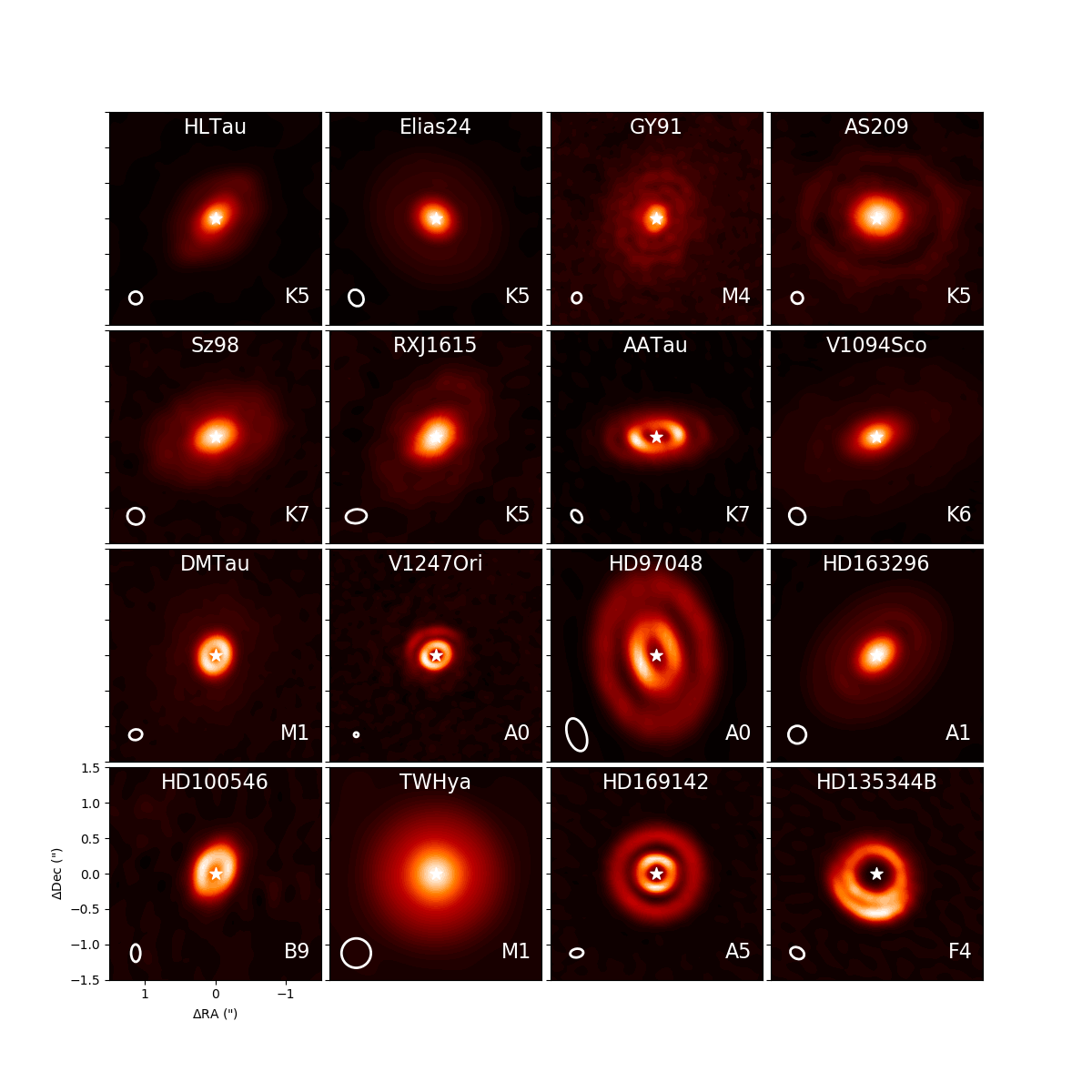}
\caption{Image gallery of the continuum emission of all disks in the sample, ordered alphabetically. The beam size is indicated in the lower left corner, the spectral type in the lower right corner. The images of HL~Tau, TW~Hya and V1247~Ori have been convolved with a 20 AU beam to be more directly comparable with the other images.}
\label{fig:originals}
\end{figure}

The ALMA continuum images have been obtained from the PIs and in a few cases from the ALMA archive. For all data, the total flux has been checked against the published integrated fluxes to make sure that the images properly recover the total integrated flux. The signal-to-noise ratio (SNR) of each image is at least 50 (Table \ref{tbl:imageprops}). Beam sizes range between 0.03" and 0.48". For the majority of the images the beam size corresponds to $\sim$20-30 AU diameter at the new Gaia distances of the targets (see Table \ref{tbl:stellar}). The images of HL~Tau, TW~Hya and V1247~Ori have a much higher spatial resolution, hampering direct comparisons. In the subsequent analysis, these three images are convolved with a 20 AU beam. Although smaller scale structures are no longer visible, this lower resolution allows a more uniform comparison for gap structure on the $\sim$20 AU scale for all disks in the sample. The only other clear exception is HD~97048, with a beam size of $>$48 AU, where even 20 AU structures remain unresolved. Considering the nature of this source with very wide gaps it is not excluded from the sample. Figure \ref{fig:originals} shows the original images of each disk, except for HL~Tau, TW~Hya and V1247~Ori, which have been convolved with a larger beam as described above. 

\begin{table}[!ht]
\label{tbl:sample}
\caption{Data sample.}
\small
\begin{tabular}{lllllllll}
\hline
Name&RA&Dec&Program ID&PI&Band&Ref&Beam(")&Beam(AU)\\
\hline
AA~Tau	&	04:34:55.427	&	+24:28:52.687	&	2013.1.01070	&	Loomis	&	6	&	1	&	0.20$\times$0.13	&	27$\times$18	\\
AS~209	&	16:49:15.295	&	-14:22:09.072	&	2015.1.00486	&	Fedele	&	6	&	2	&	0.17$\times$0.16	&	21$\times$19	\\
DM~Tau	&	04:33:48.748	&	+18:10:09.661	&	2015.1.00678	&	Qi	&	6	&	-	&	0.18$\times$0.15	&	26$\times$22	\\
Elias~24	&	16:26:24.078	&	-24:16:13.843	&	2013.1.00498	&	Perez,L.	&	6	&	3	&	0.24$\times$0.20	&	33$\times$27	\\
GY~91	&	16:26:40.470	&	-24:27:14.977	&	2015.1.00761	&	Sheehan	&	7	&	4	&	0.15$\times$0.13	&	21$\times$18	\\
HD~100546	&	11:33:25.316	&	-70:11:41.265	&	2016.1.00344	&	Perez,S.	&	6	&	-	&	0.25$\times$0.13	&	28$\times$14	\\
HD~135344B	&	15:15:48.42	&	-37:09:16.36	&	2012.1.00158	&	van Dishoeck	&	7	&	5	&	0.21$\times$0.16	&	29$\times$22	\\
HD~163296	&	17:56:21.279	&	-21:57:22.46	&	2013.1.00601	&	Isella	&	6	&	6	&	0.25$\times$0.25	&	25$\times$25	\\
HD~169142	&	18:24:29.777	&	-29:46:49.927	&	2012.1.00799	&	Honda	&	7	&	7	&	0.19$\times$0.12	&	22$\times$14	\\
HD~97048	&	11:08:03.185	&	-77:39:17.48	&	2013.1.00658	&	van der Plas	&	7	&	8	&	0.48$\times$0.26	&	89$\times$48	\\
HL~Tau	&	04:31:38.426	&	+18:13:57.238	&	2011.0.00015	&	-	&	6	&	9	&	0.035$\times$0.022	&	4.9$\times$3.1$^1$	\\
RXJ~1615.3-3255	&	16:15:20.225	&	-32:55:05.35	&	2011.0.00724	&	Perez	&	9	&	10	&	0.30$\times$0.20	&	47$\times$32	\\
Sz~98	&	16:08:22.481	&	-39:04:46.870	&	2015.1.00222	&	Williams	&	6	&	11	&	0.24$\times$0.23	&	37$\times$36	\\
TW~Hya	&	11:01:51.819	&	-34:42:17.243	&	2015.1.00686	&	Andrews	&	7	&	12	&	0.03$\times$0.03	&	1.8$\times$1.8$^1$	\\
V1094~Sco	&	16:08:36.147	&	-39:23:02.820	&	2016.1.01239	&	van Terwisga	&	6	&	13	&	0.25$\times$0.22	&	39$\times$34	\\
V1247~Ori	&	05:38:05.252	&	-01:15:21.72	&	2015.1.00986	&	Fukagawa	&	7	&	14	&	0.043$\times$0.028	&	17$\times$11$^1$	\\
\hline
\end{tabular}\\
{\bf Refs.} 1. \citet{Loomis2017}, 2. \citet{Fedele2018}, 3. \citet{diPierro2018}, 4. \citet{SheehanEisner2018}, 5. \citet{vanderMarel2016-spirals}, 6. \citet{Isella2016}, 7. \citet{Bertrang2018}, 8. \citet{vanderPlas2017}, 9. \citet{HLTau2015}, 10. \citet{vanderMarel2015-12co}, 11. \citet{Ansdell2018}, 12. \citet{Andrews2016}, 13. \citet{vanTerwisga2018}, \citet{Kraus2017}\\
$^1$ The image was reconvolved with a 20 AU beam to match the resolution of the other disks in the sample. 
\end{table}

\begin{table}[!ht]
\centering
\caption{Stellar properties}
\label{tbl:stellar}
\begin{tabular}{lllllllllll}
\hline
Name	&	SpT	&	$T_{\rm eff}$	&	$d$	&	$L_*$	&	$A_V$	&	Age (Siess)	&	$M_*$ (Siess) & Age (Bar.)	&	$M_*$ (Bar.)&	Ref.	\\
	&		&	(K)	&	(pc)	&	($L_{\odot}$)	&	(mag)	&	(Myr)	&	($M_{\odot}$) & (Myr)	&	($M_{\odot}$)&		SpT\\
\hline
AA~Tau	&	K5	&	4250	&	137$\pm$2	&	0.56$^{+0.2}_{-0.2}$	&	1.5	&	1.1-6.4	&	0.5-0.8	&	3.8-17	&	1.0-1.1	&	1	\\
AS~209	&	K5	&	4250	&	121$\pm$1	&	1.8$^{+0.2}_{-0.15}$	&	1.4	&	0.6-1.1	&	0.6-0.9	&	1.0-1.9	&	1.3-1.4	&	2	\\
DM~Tau	&	M1	&	3700	&	145$\pm$1	&	0.2$^{+0.03}_{-0.02}$	&	0.03	&	2.0-4.0	&	0.3-0.4	&	3.8-16	&	0.6-0.8	&	1	\\
Elias~24	&	K5	&	4250	&	136$\pm$2	&	6.8$^{+5.8}_{-3.2}$	&	12	&	$<$0.5	&	0.8-0.9	&	$<$0.5	&	-	&	3	\\
GY~91	&	M4	&	3300	&	137$^a$	&	0.33$^{+0.2}_{-0.2}$$^b$	&	$\sim$30$^c$	&	0.1-2.6	&	0.18-0.20	&	$<$1.5	&	0.3-0.4	&	4	\\
HD~100546	&	B9	&	10700	&	110$\pm$1	&	36.6$^{+3.4}_{-6.6}$	&	0.4	&	4.1-16	&	2.3-2.5	&	-	&	-	&	5	\\
HD~135344B	&	F4	&	6590	&	136$\pm$2	&	6.7$^{+1.3}_{-2.9}$	&	0.3	&	8.6-15	&	1.4-1.6	&	-	&	-	&	6	\\
HD~163296	&	A1	&	9250	&	102$\pm$2	&	22.7$^{+8.3}_{-7.7}$	&	1.4	&	5.0-8.9	&	1.9-2.1	&	-	&	-	&	7	\\
HD~169142	&	A5	&	8400	&	114$\pm$1	&	14$^{+3.5}_{-5.0}$	&	3.2	&	6.9-18	&	1.7-1.8	&	-	&	-	&	8	\\
HD~97048	&	A0	&	10000	&	185$\pm$1	&	31.2$^{+2.1}_{-6.0}$	&	0.9	&	4.7-6.1	&	2.1-2.3	&	-	&	-	&	9	\\
HL~Tau	&	K5	&	4400	&	147.3$\pm$0.5$^d$&	9.25$^{+5.5}_{-5.5}$$^e$	& $\sim$15$^c$&	$<$0.56	&	$<$0.6	&	1.0-1.1	&	$<$0.5	&	10	\\
RXJ~1615	&	K7	&	4060	&	158$\pm$1	&	0.83$^{+0.07}_{-0.07}$	&	0.4	&	0.9-2.0	&	0.5-0.7	&	3.0-6.0	&	1.1-1.2	&	11	\\
Sz~98	&	K7	&	4060	&	156$\pm$1	&	1.1$^{+0.2}_{-0.3}$	&	2.6	&	0.7-1.8	&	0.5-0.7	&	1.7-5.6	&	1.1-1.2	&	12	\\
TW~Hya	&	M0.5	&	3800	&	60.1$\pm$0.1	&	0.27$^{+0.06}_{-0.07}$	&	0	&	1.7-5.2	&	0.3-0.5	&	5.1-20	&	0.7-0.8	&	13	\\
V1094~Sco	&	K6	&	4200	&	154$\pm$1	&	0.57$^{+0.07}_{-0.13}$	&	2.6	&	2.0-7.0	&	0.7-0.9	&	6.0-18	&	1.0-1.1	&	12	\\
V1247~Ori	&	F0	&	7200	&	398$\pm$10	&	20.0$^{+2.7}_{-5.3}$	&	1	&	4.3-6.7	&	1.8-2.1	&	-	&	-	&	14	\\
\hline
\end{tabular}\\
{\bf Refs.} 1. \citet{Herbig1977}, 2. \citet{HerbigBell1988}, 3. \citet{Wilking2005}, 4. \citet{Doppman2005}, 5. \citet{Houk1975}, 6. \citet{Dunkin1997}, 7. \citet{Houk1988}, 8. \citet{Blondel2006}, 9. \citet{Whittet1997}, 10. \citet{White2004}, 11. \citet{Manara2014}	, 12. \citet{Alcala2017}, 13. \citet{Sokal2018}, 14. \citet{Vieira2003}\\ 
$^a$ Average distance to Ophiuchus, \citep{OrtizLeon2017}\\
$^b$ A stellar luminosity of 0.11 $L_{\odot}$ was derived by \citet{SheehanEisner2018} using a disk+cloud model, but this fit was found to be unreliable afterwards (Sheehan, private communication). We adopt a new value of 0.33 $L_{\odot}$ which appears to be more consistent with the NIR flux and the embedded nature of this source.\\
$^c$ The $A_v$ determined for an embedded object is expected to be inaccurate, so the stellar properties remain highly uncertain.
$^d$ Using the VLBA derived distance \citep{Galli2018}\\
$^e$ Derived by \citet{Robitaille2007} using a disk+cloud model\\

\end{table}

\begin{table}[!ht]
\centering
\caption{Continuum image properties}
\label{tbl:imageprops}
\begin{tabular}{lllllll}
\hline
	&	PA	&	$i$	&	Wavelength	&	Flux	&	$M_{\rm disk}$	& S/N\\
	&	($^{\circ}$)	&	($^{\circ}$)	&	(mm)	&	(mJy)	&	(M$_{\rm Jup}$)	&\\
\hline
AA~Tau	&	93	&	59.1	&	1.1	&	105	&	17	&148\\
AS~209	&	86	&	35.3	&	1.3	&	227	&	29	&74\\
DM~Tau	&	-10	&	40	&	1.1	&	109	&	20	&139\\
Elias~24	&	46.7	&	28.5	&	1.3	&	331	&	53	& 293\\
GY~91	&	-19	&	39	&	0.86	&	258	&	20	&57\\
HD~100546	&	138	&	47	&	1.3	&	379	&	40	& 90\\
HD~135344B	&	4.5	&	41.4	&	0.85	&	564	&	43	&126\\
HD~163296	&	132	&	42	&	1.3	&	721	&	65	& 317\\
HD~169142	&	62	&	16	&	0.89	&	496	&	26	&160\\
HD~97048	&	5	&	13	&	0.9	&	2253$^a$	&	315	& 820\\
HL~Tau	&	146	&	44	&	1.3	&	789	&	134	& 606\\
RXJ~1615.3-3255	&	153	&	45	&	0.44	&	878	&	44$^b$	&70\\
Sz~98	&	111.6	&	47.1	&	1.3	&	105	&	22	&106\\
TW~Hya	&	155	&	7	&	0.87	&	1495	&	22	& 60\\
V1094~Sco	&	109	&	53	&	1.3	&	204	&	42	&181\\
V1247~Ori	&	25.4	&	15	&	0.85	&	314	&	203	& 54\\
\hline
\end{tabular}\\
$^a$ Total flux reported in \citet{vanderPlas2017}, while the used image was created using superuniform weighting which only recovered about one-third of the flux.\\
$^b$ Disk mass derived from the 870 $\mu$m flux reported in \citet{Andrews2011}.\\
\end{table}

\subsection{Distances and ages}

\begin{figure}
\centering
\includegraphics[width=0.7\textwidth]{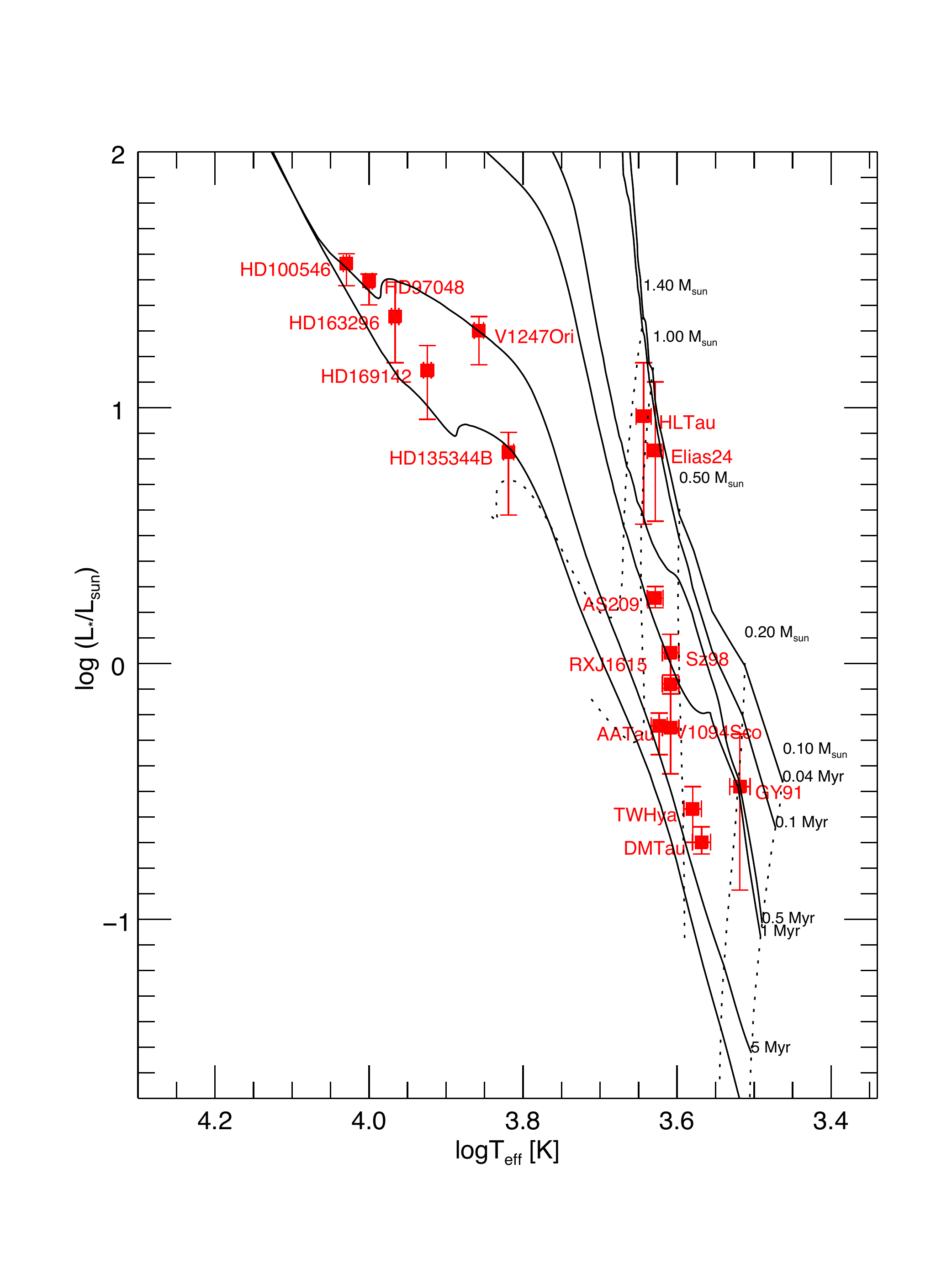}
\caption{Evolutionary diagram of the targets in this sample, with the stellar tracks from \citet{Siess2000} for a range of ages and stellar masses overplotted.}
\label{fig:HRD}
\end{figure}

\begin{figure}[!ht]
\centering
\includegraphics[width=0.8\textwidth]{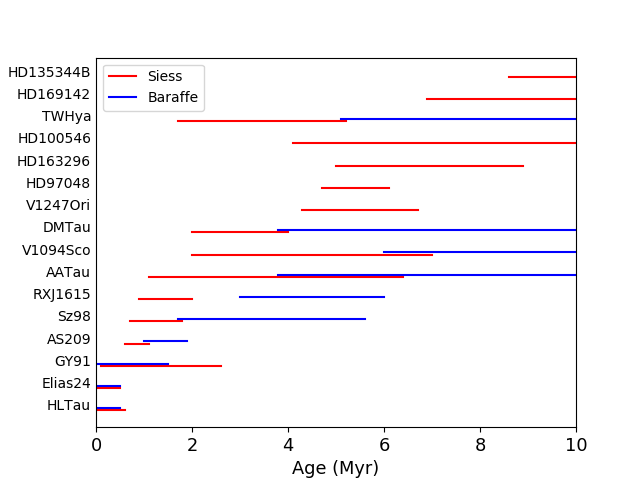}
\caption{Age ranges of the targets in the sample for each evolutionary model, sorted by age.}
\label{fig:ages}
\end{figure}

Table \ref{tbl:stellar} lists the stellar properties and ages for each target. Spectral types have been taken from the literature.  As accretion was derived using different methods for each target, we have not included this particular property in the table. For all targets optical spectroscopy was used to determine the spectral type, except for GY~91 which is optically extincted, and spectral typing was performed using near infrared spectroscopy. Distances have been taken from the \emph{Gaia} Data Release 2 \citep{Gaia2018}. HL~Tau and GY~91 are not available in the \emph{Gaia} catalog due to their nebulosity, so we use instead the VLBA derived distance for HL~Tau of 147.3 $\pm$0.5 pc \citep{Galli2018} and the average cloud distance of Ophiuchus \citep[137 pc,][]{OrtizLeon2017}, respectively. Near infrared photometry data (0.8-2.2 $\mu$m) were fit to Kurucz models \citep{CastelliKurucz2004} to derive the stellar luminosities (and uncertainties) and optical extinctions $A_V$ using $R_V$=3.1 for all targets considering the new \emph{Gaia} distances. For the embedded objects (GY~91 and HL~Tau), it became clear that a simple extinction law is not applicable to fit the extinction due to the ambient cloud. For HL~Tau we use the results from a more detailed modeling procedure of a disk+envelope model from the literature \citep{Robitaille2007}. For GY~91, \citet{SheehanEisner2018} find $L_*$=0.11 $L_{\odot}$ for a disk+cloud model, but this luminosity is below expectations for a deeply embedded M4 star. Our photometric fitting indicates a somewhat larger value of $L_*\sim0.33 L_{\odot}$, which is adopted in our analysis (albeit with a large error bar due to the large uncertainties in the dereddening of an embedded object).

Individual stellar ages are derived using the new luminosities by interpolating the evolutionary tracks of both \citet{Siess2000} and \citet{Baraffe2015} (Figure \ref{fig:HRD}). An uncertainty of 100 K is adopted for the effective temperature. The Baraffe et al. models are only applicable for stars $<$1.4 $M_{\odot}$ and ages $>$0.5 Myr, and thus are not suitable for all the stars in our sample. Figure \ref{fig:HRD} shows the Siess evolutionary tracks and our targets. The results from the two evolutionary models generally agree (where both available), although both TW~Hya and DM~Tau are significantly older according to the Baraffe et al. models. Age determination of individual stars using this method remains highly uncertain and extreme care has to be taken when using these numbers, which is the reason that we merely report age and mass ranges. \citet{Sokal2018} analyzed the spectral type of TW~Hya in high detail, and found an age of 8$\pm$3 Myr and a mass of 0.6$\pm$0.1 $M_{\odot}$, which is within our derived estimates using a simpler approach. Despite the large uncertainties in age determination, these numbers do provide a way to order targets by approximate age, distinguishing between younger, intermediate and older disks (Figure \ref{fig:ages}). In the subsequent analysis, the targets are ordered by their approximate age following these results.

\clearpage
\newpage

\section{Analysis}
\begin{figure}[!ht]
\centering
\includegraphics[width=\textwidth]{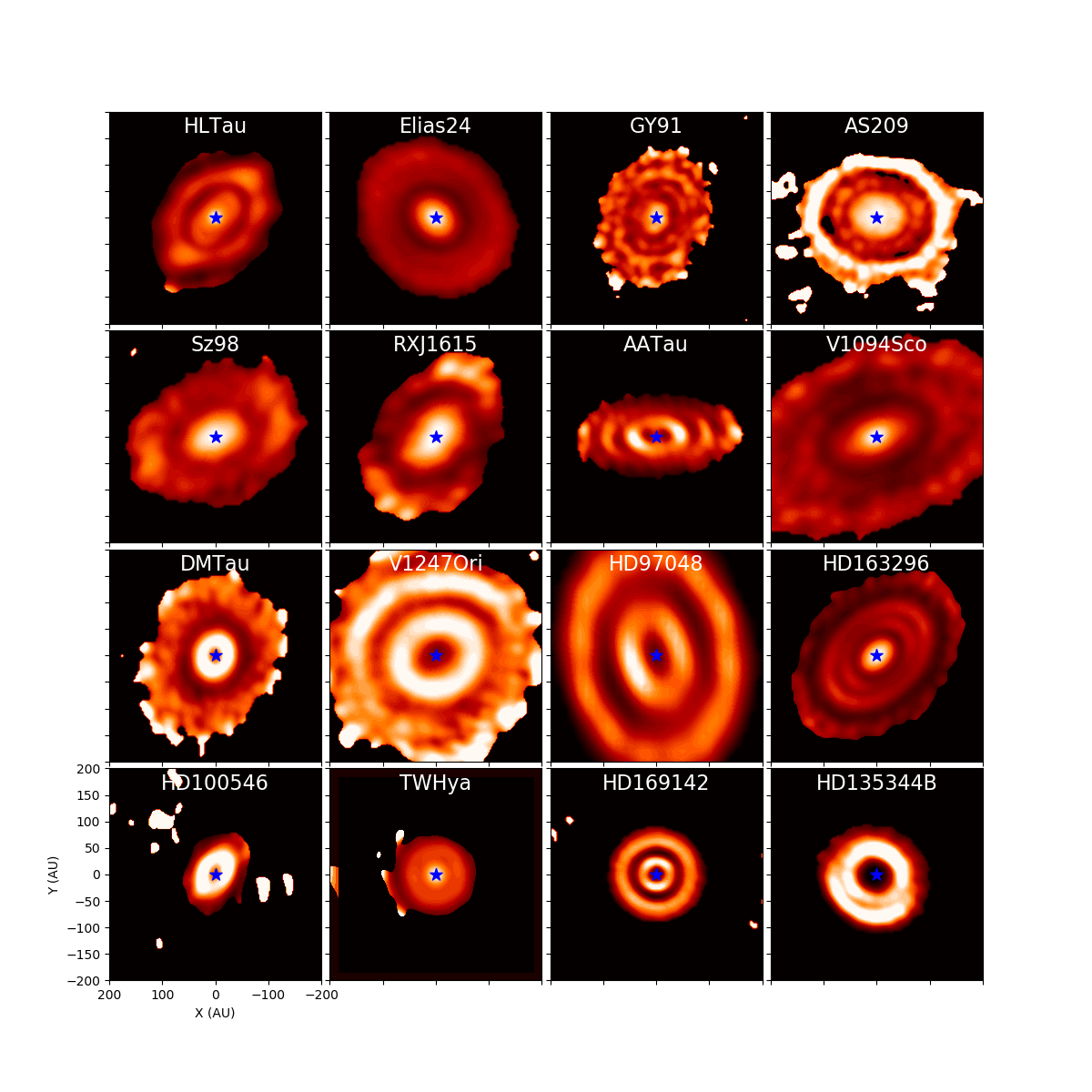}
\caption{Image gallery of the continuum emission of all disks in the sample in the EIR representation on the same spatial scale (400$\times$400 AU). The disks are ordered by age.}
\label{fig:overview_EIR}
\end{figure}

In this study, we aim to analyze the radial locations and widths of the major gaps for all disks in the sample. Narrower gaps (such as those seen in the much higher resolution images of HL~Tau and TW~Hya) are ignored. The analysis is performed in the image plane which is sufficient for derivation of gap locations. For the analysis, the images are first produced using  the enhanced imaging representation \citep[EIR,][]{vanderMarel2018-hd16}, a variant of unsharp masking. In this technique, the normalized image is convolved with a beam 2 times larger than the original beam, and the original is divided by this convolved image. The resulting ratio image shows more clearly the locations of the gaps and edges. Furthermore, the images are scaled to their distances. Figure \ref{fig:overview_EIR} shows an overview of the disks in EIR at the same AU spatial scale, ordered by age.

The gaps are analyzed using the azimuthally averaged intensity profiles of the EIR images, after deprojection and normalization to the peak. Inclinations and position angles are taken from the literature and listed in Table \ref{tbl:imageprops}. This Table also lists the total millimeter flux and the corresponding disk mass, using the relations presented in \citet{Cieza2008}, following \citet{AndrewsWilliams2007oph}, assuming a disk temperature of 20 K, a dust opacity $\kappa_{\nu}$=10 cm$^2$ g$^{-1}$ at 1000 GHz \citep{Beckwith1991}, a dust opacity law coefficient $\beta$=1 (considering $\kappa_{\nu}\sim\nu^{\beta}$) and a gas-to-dust ratio of 100. 

\begin{figure}[!ht]
\centering
\includegraphics[width=\textwidth]{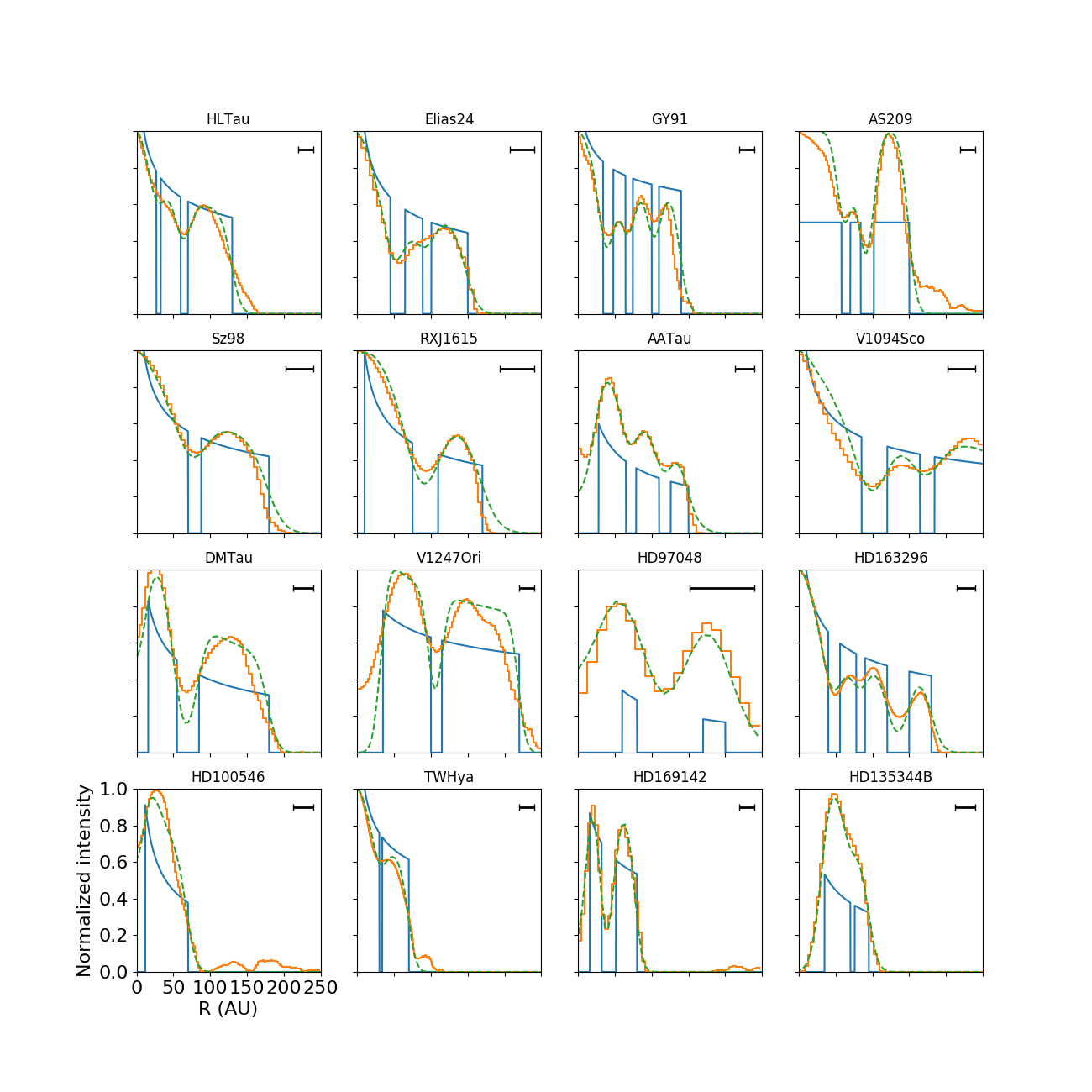}
\caption{Azimuthally averaged and normalized intensity profiles of the EIR images of Figure \ref{fig:overview_EIR} of the disks in the sample. The orange profiles represent the data, the green profiles are the best-fit models (convolved with the beam) and the blue profiles the best-fit models without convolution. The model is a simple power-law with empty gaps. The parameters are given in Table \ref{tbl:fitparameters}. The beam size is indicated in the upper right corner.}
\label{fig:azimcuts}
\end{figure}

\begin{figure}[!ht]
\centering
\includegraphics[width=\textwidth]{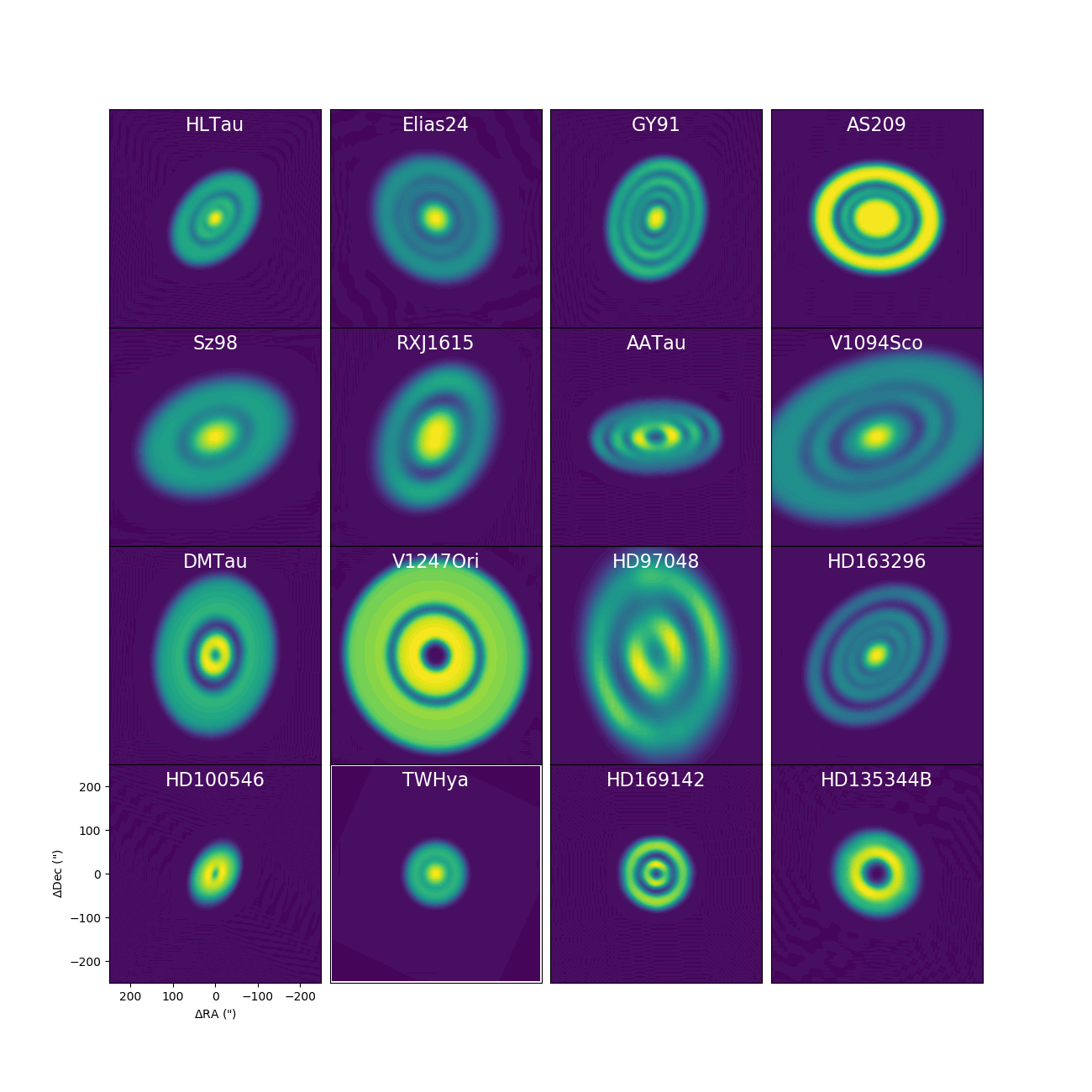}
\caption{Best fits of the EIR images of Figure \ref{fig:overview_EIR} of the disks in the sample, all on the same spatial scale (400$\times$400 AU). The disks are ordered by age. The intensity profiles are normalized, just like the EIR images.}
\label{fig:imageresults}
\end{figure}

The intensity profiles and best fits are shown in Figure \ref{fig:azimcuts} and the images with residuals in Figure \ref{fig:imageresults}. The intensity profile is fit using a $\chi^2$ minimization with a power-law profile $I(r) = (r/r_c)^{-\gamma}$, with $r_c$=1 and $\gamma$ a free parameter between 0.0 and 1.0. Gaps are implemented by setting $I(r)$ equal to zero between $r_x$ and $r_{xw}$, for $x$=1,2,3... depending on the number of (visible) gaps. For the fit, an initial guess of each gap center and width is chosen based on visual inspection and varied in steps of 1 AU for the $\chi^2$ minimization. In several cases, additional gaps were added after the first round of fitting to obtain a better match with the images and their width and center were again varied in steps of 1 AU. The number of free parameters ranges from 4 to 8 depending on the number of gaps. The intensity profile is convolved with the beam size for comparison with the data. Best-fit parameters are listed in Table \ref{tbl:fitparameters}. Gaps are described as values ranging between $r_x$ and $r_x+r_{xw}$. Clearly this method does not account for depth of the gap or absolute intensity variations, it is only a way to quantify the gap locations. 

\begin{figure}[!ht]
\centering
\includegraphics[width=0.6\textwidth]{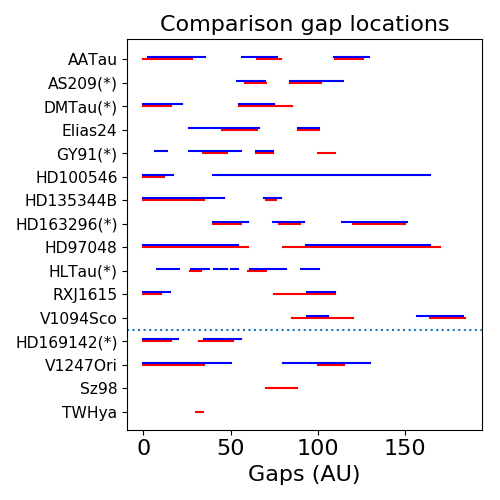}
\caption{Comparison between our derived gaps (red) and the derived gaps of the individual studies (blue) in the literature (Table \ref{tbl:sample}), after scaling to the new Gaia distance. The targets above the dashed line were modeled in the visibility plane, the ones below in the image plane. Targets with an asterisk (*) were modeled assuming a surface density depletion rather than an empty gap. For Sz~98 and TW~Hya, no fitting of the gap locations was performed in the literature and no comparison is possible. Considering the different modeling approaches in the literature, our gap locations generally agree well with previous results.}
\label{fig:comparison}
\end{figure}

Although our derivation of gap locations and widths in the image plane is simple, we find a good correspondence between our parameters and those derived in the literature where the data was first analyzed (Figure \ref{fig:comparison}). The relevant papers are listed in Table \ref{tbl:sample}). The DM~Tau data was not published before, but \citet{Kudo2018} derives drops in the surface density based on a newer, proprietary dataset, which is used in our comparison. For HL~Tau and HD~169142, no gap parameters were derived in the original papers, so we use values from follow-up papers, \citet{Pinte2016} and \citet{Fedele2017}, for these two disks respectively. For Sz~98 and TW~Hya, no gap fitting has been performed before and no comparison could be made.

Most of these studies fit the continuum in the visibility plane. Some authors have chosen to do intensity modeling using either an approach with a power-law profile and gaps or a superposition of gaussian rings, whereas others run a full radiative transfer with the aim of deriving the dust surface density and the amount of depletion of dust inside the gaps (indicated by a star in Figure \ref{fig:comparison}). Due to the diversity in modeling approaches, the results are expected to vary throughout the sample and compared with our results, but the gap locations and widths are generally consistent with our findings.

For GY~91, the inner gap around 10 AU was not recovered in our modeling. This gap was derived by \citet{SheehanEisner2018} from the 100 GHz data with much higher spatial resolution than the 345 GHz data. The outer gap around 110 AU that we derive is visible in \citet{SheehanEisner2018} but was not included in the fit, perhaps due to the low signal to noise. For HD~100546, \citet{Walsh2014} reports an additional outer ring at 150-225 AU at $<$1\% of the peak level. Considering the SNR and spatial filtering of the presented data, this ring is not recovered and their is no counterpart of the outer gap. For both HD~135344B and V1247~Ori, the images show an asymmetry rather than a ring, and our `gap' fitting assuming an azimuthally averaged ring naturally results in a narrower outer gap. Still, fitting the azimuthally averaged profile is a reasonable way to describe the gap location, if it is assumed that the asymmetry is the result of azimuthal trapping within a dust ring. For HL~Tau, only two of the largest gaps are recovered in our analysis due to the convolution with the larger beam size. Also, several of the other gaps derived by \citet{Pinte2016} are less than a factor 10 depleted in dust surface density which cannot be recovered in our approach with empty dust gaps.

\begin{table}[!ht]
\centering
\caption{Fit parameters}
\label{tbl:fitparameters}
\begin{tabular}{llllll}
\hline
Name&$\gamma$&Gap 1&Gap 2&Gap 3&$R_{\rm out}$\\
&&(AU)&(AU)&(AU)&(AU)\\
\hline
AA~Tau	&	0.5	&	0-28	&	65-79	&	110-126	&	150	\\
AS~209	&	0	&	58-70	&	84-102	&		&	150	\\
DM~Tau	&	0.4	&	0-16	&	55-85	&		&	180	\\
Elias~24	&	0.3	&	45-65	&	88-100	&		&	150	\\
GY~91	&	0.15	&	34-48	&	64-74	&	100-114	&	140	\\
HD~100546	&	0.5	&	0-12	&		&		&	70	\\
HD~135344B	&	0.5	&	0-35	&	70-76	&		&	95	\\
HD~163296	&	0.3	&	40-56	&	78-90	&	120-150	&	180	\\
HD~169142	&	0.3	&	0-16	&	32-52	&		&	80	\\
HD~97048	&	0.6	&	0-60	&	80-170	&		&	200	\\
HL~Tau	&	0.25	&	27-33	&	60-70	&		&	120	\\
RXJ~1615.3-3255	&	0.35	&	0-10	&	75-110	&		&	170	\\
Sz~98	&	0.3	&	70-88	&		&		&	180	\\
TW~Hya	&	0.25	&	30-34	&		&		&	70	\\
V1094~Sco	&	0.3	&	55-65	&	85-120	&	164-184	&	290	\\
V1247~Ori	&	0.2	&	0-35	&	100-115	&		&	220	\\
\hline
\end{tabular}
\end{table}

\clearpage
\newpage

\section{Discussion}
\subsection{Origin of the gaps}
\begin{figure}[!ht]
\includegraphics[width=0.45\textwidth]{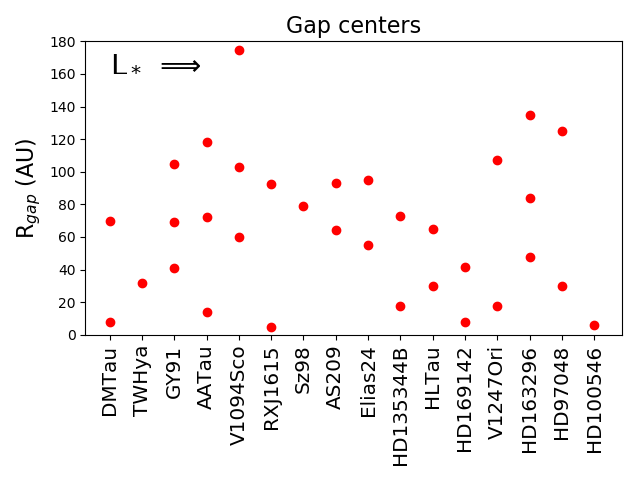}
\includegraphics[width=0.45\textwidth]{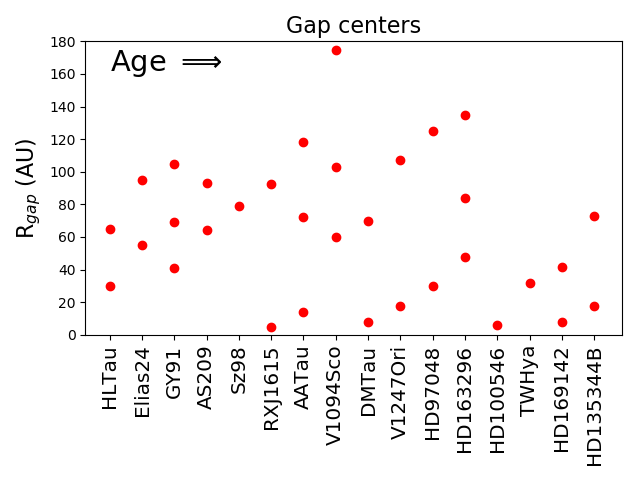}\\
\includegraphics[width=0.45\textwidth]{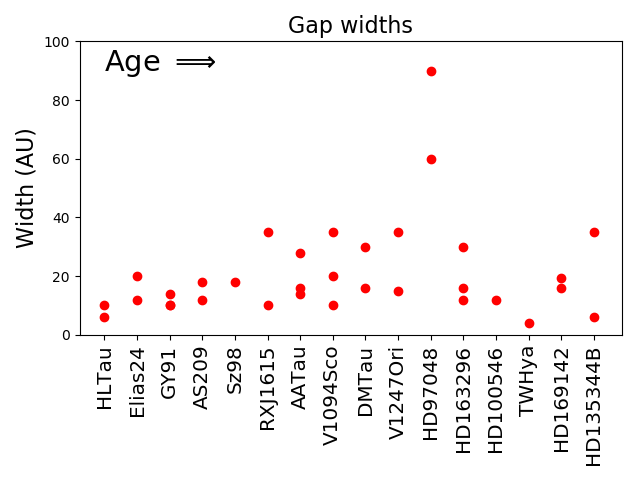}
\includegraphics[width=0.45\textwidth]{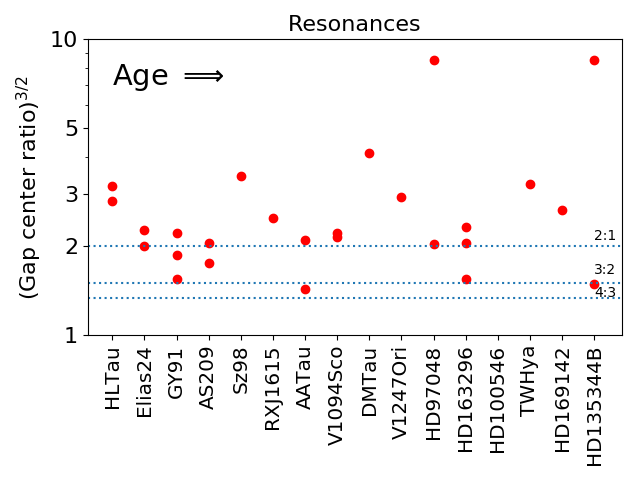}\\
\includegraphics[width=0.45\textwidth]{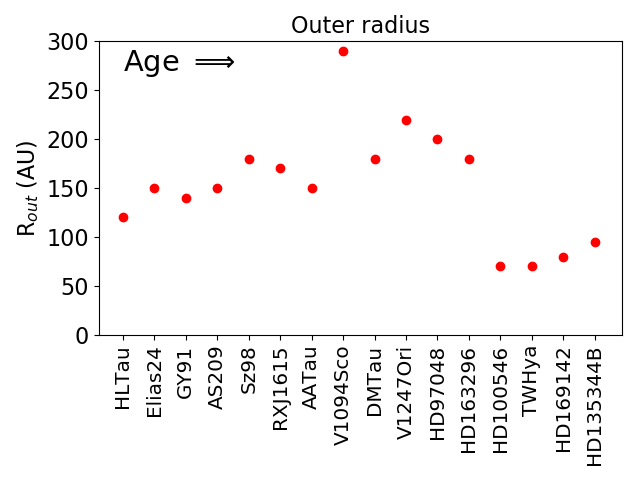}
\caption{Trends based on the derived fit parameters (Table \ref{tbl:fitparameters}) of the gap properties in each disk. The first plot shows the gap centers sorted by increasing stellar luminosity, the other plots are sorted by increasing age. No obvious trends are visible in any of the plots showing gap properties. The outer radius drops off for the older disks.}
\label{fig:trends}
\end{figure}

With the derived properties, it is possible to look for trends in the disks within the sample. It is clear that all disks show prominent gaps (or rings, as an inverse description), regardless of age or stellar properties. In Figure \ref{fig:trends} the derived parameters are plotted as a function of age and luminosity. The chosen parameters are the gap center radii, gap widths and the relative gap distances (the ratio between the central radii of the subsequent gaps with the first gap radius), the outer radius and the $\gamma$ value of the power-law. There is a bias in the sample due to the lack of young, massive Herbig stars and (except for TW~Hya) a lack of older low-mass T Tauri stars, which needs to be taken into account when interpreting these plots. 

No obvious trends are seen in Figure \ref{fig:trends}. For the gap centers, widths and even ratios between gap centers no trends are visible, which makes a derivation of a common origin challenging. The only clear trend seen is in the outer radius, which decreases suddenly for the oldest disks in the sample, which is discussed further in Section \ref{sect:outer}. 

\subsubsection{Snow lines}
\begin{figure}[!ht]
\centering
\includegraphics[width=\textwidth]{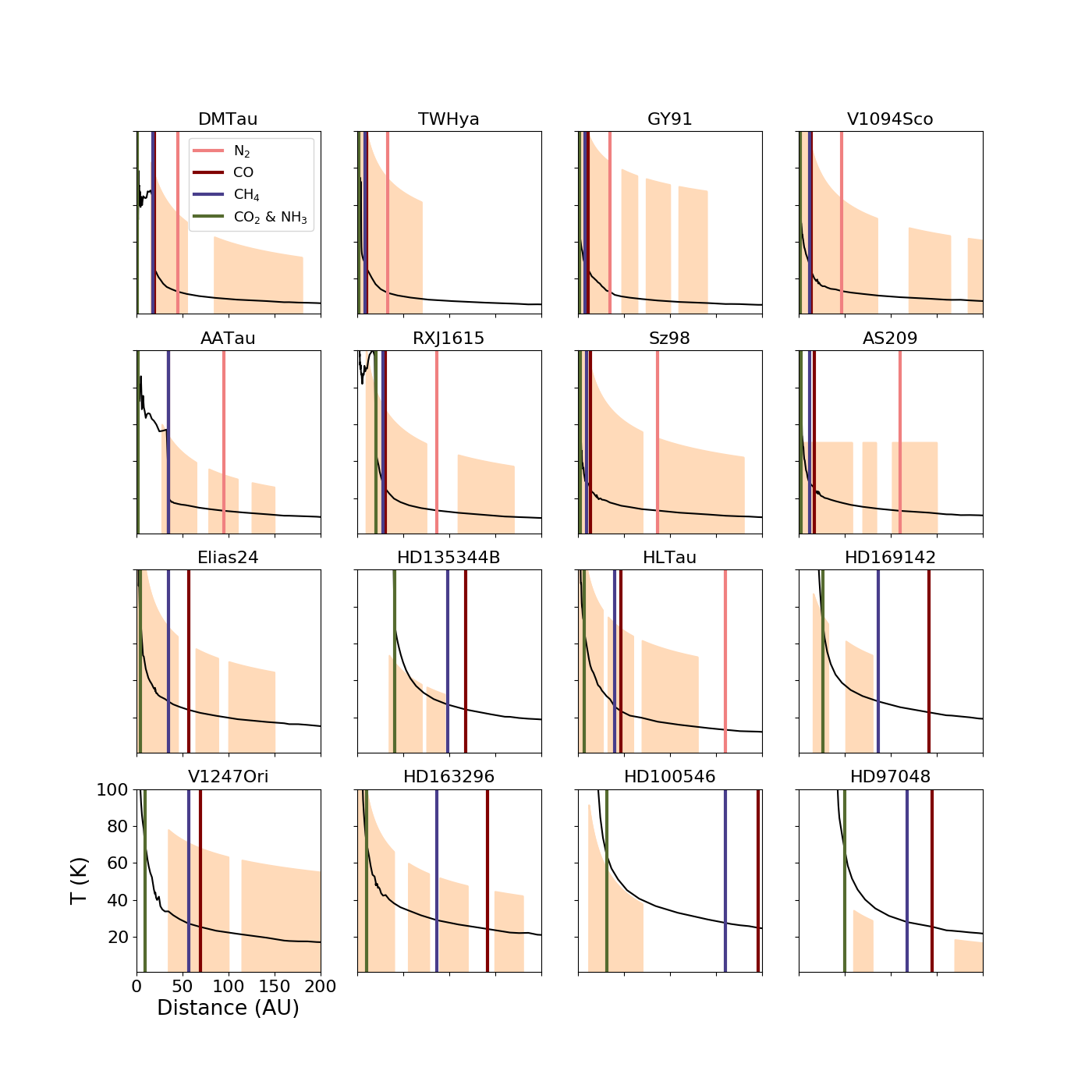}
\caption{Dust midplane temperature profiles of the radiative transfer models given in Figure \ref{fig:SEDs} in black lines. Targets are sorted by luminosity. The radii of the main freeze out temperatures considered here are indicated by vertical solid lines (N$_2$ in salmon, CO in brown, CH$_4$ in blue and the mean value of CO$_2$ and NH$_3$ (which are close together) in green). The derived intensity profiles with the gaps are shown in orange. Clearly, there is no clear correlation between the locations of the gaps and the snow lines.}
\label{fig:temperatures}
\end{figure}

The lack of correlation between stellar luminosity and gap location (Figure \ref{fig:trends}) puts the snow line mechanism  \citep{Zhang2015} into question. To check this mechanism, we have run a dust radiative transfer model for each disk, using the stellar properties in Table \ref{tbl:stellar}, with the code DALI \citep{Bruderer2013} following the parametrization as described in \citet{Andrews2011}. The vertical parameters and surface density are adjusted to be consistent with the SED and the millimeter flux. For simplicity, only the inner dust cavities are included in the surface density model where applicable, not the gaps further out, as outer gaps do not leave signatures in the SED slope. The SEDs are shown in Figure \ref{fig:SEDs}.

The midplane dust temperature of each radiative transfer model provides a way to locate the approximate radii where a snow line is expected. The dust temperature profiles are presented in Figure \ref{fig:temperatures}, and we consider the average freeze-out temperatures of 5 molecules: N$_2$ (13.5 K), CO (25.5 K), CH$_4$ (29 K), CO$_2$ (66 K) and NH$_3$ (80 K), following \citet{Zhang2015} and \citet{Okuzumi2016}. The comparison of the gap locations with the radii of these freeze-out temperatures reveals that even though there is overlap in some cases, there is clearly no general trend, and snow lines cannot be considered as a main mechanism for the origin of rings and gaps in protoplanetary disks. A similar conclusion was recently found for a number of targets in the Taurus star forming region \citep{Long2018} and the DSHARP program \citep{Huang2018} for a narrower range in spectral types and age. Although there are uncertainties in the parameters used in radiative transfer which could lead to changes in the temperature profile, it is unlikely that this can shift \emph{all} gaps to the proper locations: systematic shifts and ratios would be visible. A link between stellar luminosity (corresponding to the stellar radiation field) and the gap locations is missing. If the luminosity is variable with time due to stellar mass build up and episodic accretion or outbursts such as in FU Orionis objects, snow lines may temporarily move outwards in the disk \citep[e.g.][]{Cieza2016}. The time scales of these variations are relatively short, however, and cannot explain the large number of gaps in our sample that are inconsistent with freeze out radii. This large sample approach thus provides evidence that the snow line scenario is not able to explain the majority of disks with rings. 

\subsubsection{Planets}
Planets clearing gaps are often invoked as a possible explanation for gaps in disks and our results do not provide strong evidence to either support or reject this conclusion. The deep, wide gaps at tens of AU indicate planets with masses of the order of $M_{\rm Jup}$, but Jovian planet formation is challenging at large orbital radii due to the required time scales for core accretion and required disk masses for gravitational instability \citep{Helled2014}. Whereas exoplanet statistics suggest that Jovian planets are uncommon within the assumptions of the hot-start model \citep{Bowler2016}, recent results show that when the cold-start model is assumed instead, the occurrence rate of giant planets at wide orbits can be as high as 90\% \citep{Stone2018}. (Sub)-Neptune planets are capable of carving gaps in disks if the disk viscosity is low enough \citep[e.g.][]{Rosotti2016,Dipierro2017,Bae2017,Dong2017multi,Dong2018gaps}. The low turbulence measured in a handful of disks \citep{Flaherty2015,Teague2016,Flaherty2018} suggests that viscosity may be lower than expected, in favor of this scenario. 

Direct imaging searches for planets in ring systems have so far remained unsuccessful. No signs of planets were found in deep searches in HL~Tau \citep{Testi2015} and only a tentative detection of a point source in HD~163296 \citep{Guidi2018}, while searches for giant planets in transition disks have mostly remained unsuccessful \citep[e.g.][]{Maire2017}. The brightness estimates of young planets, however, are subject to the question of whether their formation is hot-start or cold-start \citep[e.g.][]{Marley2007,Fortney2008}, so the lack of detections do not yet rule out giant planets in disks. Neptune-mass planets, which remain undetectable with current facilities, are certainly still possible and planets remain a likely scenario to explain the gaps.

\subsubsection{Gap center ratios}
There is no obvious trend in the gap center ratios which could be linked to resonances. Figure \ref{fig:trends} shows the ratios between subsequent gap centers to the power 3/2, to mimic orbital resonance ratios. Less than one third of the disks have ratios close to the first-order mean motion resonances such as 2:1 and 3:2, but many gaps fall outside this range. \citet{Boley2017} showed that the narrower gaps (unresolved in our images) in HL~Tau and TW~Hya could be reproduced by planets in resonance, but this does not appear to be a general trend for the larger scale gaps studied here. Considering the uncertainties in the unresolved data it is not possible to derive certain conclusions from this. The only visible trend is that in every disk a gap appears to be present around $\sim$50-70 AU, which is also the approximate outer radius in the older disks. This 'trend' could be a selection effect, considering the resolution of 20-30 AU. 

\subsubsection{Disk mass}
The disks in the sample are relatively massive with an average disk mass of 55 $M_{\rm Jup}$, compared to the average disk mass of $\lesssim$ 1 M$_{\rm Jup}$ in other star forming regions \citep[][]{Andrews2013,Ansdell2016,Pascucci2016}. Such high mass is likely an observational selection effect, as only the brightest disks have been imaged at high angular resolution so far. Furthermore, our sample is biased towards early-type stars compared to the typical stellar distribution in star-forming regions. Most disks, however, are well below the Toomre criterion for global gravitational instability \citep[$Q<1$,][]{Toomre1964,Boss1997}, with a typical $M_{\rm disk}/M_*$ of only $\sim5\pm3$\%. As $Q$ varies with radius, it is still possible that the outer parts of the disk follow this criterion. General gravitational instability results in spiral arms \citep{Lodato2004} which are not observed in our sample, but a secular gravitational instability (requiring gas-to-dust ratios $<$100) can generate ring-like structures in the dust \citep{Takahashi2014,Takahashi2016}. As the gas-to-dust ratio in these disks remains highly uncertain, it is not possible to exclude the possibility of secular gravitational instability as origin of the rings.

\subsection{Outer radius}
\label{sect:outer}
A prominent feature in Figure \ref{fig:trends} is the outer dust disk size, which is clearly smaller in the four oldest disks than in the younger ones. With an estimated uncertainty on individual radii of half the beam size, the mean outer radius of the 12 young disks is 178$\pm$48 AU, and for the 4 oldest disks it is 79$\pm$17 AU. Considering the high SNR of the data (Table \ref{tbl:imageprops}), this result is unlikely to be a sensitivity effect. Significant spatial filtering is excluded (except for HD~100546, as discussed above) by the original papers presenting these data. A decline of dust disk size is often associated with radial drift \citep{Brauer2008,Birnstiel2014}, but this explanation is inconsistent with the presence of the outer dust rings, which are expected to trap dust and prevent their inward drift. Namely, if the sample is revealing an evolutionary trend, it is expected that the older disks used to have outer rings as well. If no outer rings are seen in the oldest disks, this could be evidence that outer rings disappear faster than the inner disk rings. HD~100546 is an interesting example, as \citet{Walsh2014} demonstrated that a faint outer dust ring at 190 AU is present in this disk, at only 1\% of the intensity of the inner ring, based on lower resolution (but highly sensitive) observations. The observations in our sample have sufficient coverage at large angular scales but do not have the sensitivity to detect an outer ring at that level.

\begin{table}[!ht]
\begin{center}
\caption{Outer radii gas and dust}
\label{tbl:outer}
\begin{tabular}{llllll}
\hline
Name&$R_{\rm dust}$&$R_{\rm gas}$&Gas tracer&Ratio&Ref.\\
&(AU)&(AU)&&&\\
\hline
HL~Tau&120&-&-&-&-\\
Elias~24&150&-&-&-&-\\
GY~91&140&-&-&-&-\\
AS~209&150&300&$^{12}$CO 2--1&2.0&1\\
Sz~98&180&360&$^{12}$CO 2--1&2.0&2\\
RXJ1615&170&475&$^{12}$CO 3--2&2.8&3\\
AA~Tau&150&$>$220&$^{13}$CO 3--2&$>$1.3&4\\
V1094~Sco&290&438&$^{12}$CO 2--1&2.0&2\\
DM~Tau&180&1000&$^{12}$CO 3--2&5.6&5\\
V1247~Ori&220&$>$280&$^{12}$CO 3--2&$>$1.3&6\\
HD~97048&200&960&$^{12}$CO 3--2&4.8&7\\
HD~163296&180&550&$^{12}$CO 2--1&3.0&8\\
HD~100546&70&390&$^{12}$CO 3--2&5.6&9\\
TW~Hya&70&220&$^{12}$CO 3--2&3.1&10\\
HD~169142&80&195&$^{12}$CO 2--1&2.4&11\\
HD~135344B&95&205&$^{12}$CO 3--2&2.2&3\\
\hline
\end{tabular}\\
{\bf Refs.} 1. \citet{Huang2016}, 2. \citet{Ansdell2018}, 3. Archival data 2012.1.00870, 4. \citet{Loomis2017}, 5. Archival data 2013.1.00647, 6. \citet{Kraus2017}, 7. \citet{vanderPlas2017}, 8. \citet{Isella2016}, 9. \citet{Walsh2014}, 10. \citet{Huang2018twhya}, 11. \citet{Fedele2017}\\
\end{center}
\end{table}

The dispersal of gas in the outer disk could be tested with a measure of the gas outer radius. The most sensitive tracer is a $^{12}$CO low-J line. Table \ref{tbl:outer} lists gas outer radii of most of our sample based on $^{12}$CO observations as reported in the literature. For the embedded disks HL~Tau, Elias~24, and GY~91, no estimates are available as the line data are confused with the envelope emission. For AA~Tau, only $^{13}$CO observations were available, so the reported radius is a lower limit. The outer radii are likely lower limits in general, as the $^{12}$CO emission itself depends on the presence and size distribution of dust \citep{Facchini2017rout}, CO freeze out, and the detection is sensitivity-limited. This issue is particularly relevant since the targets in our sample have not been observed at the same sensitivity level. These values thus provide only a rough estimate of the size of the gas disk. The dust outer radius as derived from the intensity profile (Table \ref{tbl:fitparameters}) is given as well, and the ratio between the two. Interestingly, the decline in gas outer radius follows the same trend as the dust outer radius. Indeed, the ratios between the two radii remain more or less constant throughout the sample, suggesting that gas indeed disperses in the outer disk. A possible explanation is that the outer gas disk dissipates through external photoevaporation. Although external photoevaporation is generally associated with the presence of a nearby OB-star, which is not the case for any of our targets, recent work has revealed evidence for the importance of external photoevaporation by weak radiation fields in clusters \citep{Haworth2017,Winter2018}. Considering the large uncertainties in the derivation of the outer radius, however, a more detailed study of the gas outer radius is required.

Stellar flybys are unlikely to explain the stripping off of the outer rings, due to the lower stellar density in nearby star forming regions \citep{Megeath2016}. Specifically, stellar encounters at $\sim$60 AU radius on $<$10 Myr timescales are unlikely to be relevant. 

If the outer dust rings disappear faster than the inner dust rings, it is possible that the dust particles there have grown into planetesimal sizes, beyond observability in the millimeter regime. Such outer rings would be equivalent to Kuiper belts or debris disk rings, which are considered to be remnants of planetesimal collisions with only a fraction of the protoplanetary disk mass \citep[$\sim10^{-1}-10^{-2} M_{\rm Earth}$, e.g., ][]{Wyatt2015}. Millimeter imaging of debris disks has revealed a large number of debris disks that contain thin dust rings at large orbital radii \citep[e.g.][]{Marino2016,Booth2016,Macgregor2017}. As observed debris disks are closer, the sensitivity of these observations is significantly higher than those of the protoplanetary disk observations used in our study, so a similar kind of ring would remain undetected in our observations. A potential scenario is that the dust particles in the outer ring grow up to planetesimal sizes, leaving a small signal in the millimeter dust continuum due to continuous fragmentary collisions, to the same level as debris disks. Ultimately the inner rings disappear as well, either through growth to planets, disk dissipation or a combination of both, and only the outer planetesimal belt remains. A correlation between stellar luminosity and  planetesimal belt radius in debris disks was interpreted as a potential link with CO snow lines \citep{Matra2018}, but the lack of correlation in our study between ring radii and temperature profile directly challengdes this hypothesis. 

\subsection{Transitional disks}
Interestingly, the presence of an inner cavity seen in the transition disks of the sample, which is often considered to be related to the presence of giant planets in the inner disk, does not appear to have any correlation with the structure in the outer disk. Both double rings and triple rings are present at a range of radii regardless of the presence (or lack thereof) of a dust cavity. The only observational trend seen in our sample is the lack of inner cavities in the youngest disks in the sample, but given the small number of targets, the major interest in transition disks and the larger distances of observed embedded disks this could simply be a selection effect. Any scenario explaining the origin of transition disk cavities, however, involves time scales of a few million years \citep{Espaillat2014}, challenging the existence of transition disks of age $\lesssim$1 Myr. The young WL~17 disk \citep{SheehanEisner2017} has been suggested to be an embedded transition disk, but the high optical extinction in that system can be equally well explained by interstellar rather than circumstellar material and there is no evidence for a significant envelope emission based on the short baseline data. Overall our sample study shows clear similarities between transition disks and full disks, suggesting a common evolutionary route, as proposed by \citet{vanderMarel2018}. 

Such a connection could be envisioned as a series of consecutive events, where gaps and rings are formed early-on by some unknown mechanism, creating dust traps in various locations in the disk. Dust traps in the inner part of the disk eventually lead to the formation of a planet, which clears an inner cavity. As observed dust cavities are still very large compared to exoplanet orbital radii, these giant planets will have to migrate inwards to be consistent with exoplanet demographics \citep{vanderMarel2018}. This scenario is clearly distinct from the proposed model of triggered inside-out planet formation \citep{Chatterjee2014}, where the formation of a first planet in the inner part of the disk leads to dust traps and follow-up planet formation, when scaled up to larger orbital radii. The consistent presence of dust rings at every age of the disk regardless of an inner cavity does not support this idea for the larger spatial scales discussed here.

\section{Conclusions}
We have studied a sample of 16 disks which show multiple rings in their dust continuum emission observed with ALMA at a typical resolution of 25 AU. The targets span a wide range of ages, spectral types and luminosities, allowing a search for general trends rather than the target studies on an individual  basis that have been conducted so far. For each target, the new \emph{Gaia} distance was used to rederive the stellar luminosity and the corresponding approximate age and stellar mass using evolutionary track models. The dust continuum images were analyzed by fitting an intensity profile containing multiple gaps to each target, to derive the gap locations and widths. We find:

\begin{enumerate}
\item Disks are intrinsically diverse and there are no obvious trends in the locations of gaps and rings as a function of age that could point towards a systematic evolutionary effect.
\item There is no correlation between the locations of the snow lines of common molecules, such as CO, CO$_2$, CH$_4$, N$_2$ and NH$_3$, as derived from radiative transfer models for each individual target, and the gap locations. This result indicates that the snow line scenario cannot explain the presence of gaps and rings in disks.
\item Planets are a possibility to explain the gaps, especially if the disk viscosity is low and the gaps can be explained by sub-Neptune mass planets. 
\item The outer radius of the continuum disk decreases for the oldest disks in the sample, indicating that dust particles in the outer rings disappear faster than the inner rings. This result suggests that the outer dust rings either disappear together with the gas or grow into larger planetesimals and form planetesimal belts such as seen in debris disks and the Kuiper Belt.
\item Transitional disks with cleared inner dust cavities do not appear to have different gap structures in the outer disk, suggesting an evolutionary scenario where ring disks evolve into transitional disks. 
\end{enumerate}

  \begin{acknowledgements}
  The authors would like to thank Aaron Boley, Brenda Matthews and Sam Lawler for useful discussions and the referee for their thoughtful comments. We would also like to thank Sierk van Terwisga, Gerrit van der Plas, Stefan Kraus and Takayuki Muto for providing the reduced ALMA data on V1094 Sco, HD~97048 and V1247~Ori, respectively.
This paper makes use of the
  following ALMA data: ADS/
JAO.ALMA\#2011.0.00015.SV, 2011.0.00724.S, 2012.1.00158.S, 2012.1.00799.S, 2013.1.00498.S, 2013.1.00601.S, 2013.1.00658.S, 2013.1.01070.S, 2015.1.00486.S, 2015.1.00678.S, 2015.1.00761.S, 2015.1.00222.S,2015.1.00686.S, 2015.1.00986.S, 2016.1.00344.S, 2016.1.01239.S. ALMA is a partnership of ESO (representing its member states), NSF (USA) and
  NINS (Japan), together with NRC (Canada) and NSC and ASIAA (Taiwan),
  in cooperation with the Republic of Chile. The Joint ALMA
  Observatory is operated by ESO, AUI/NRAO and NAOJ. Part of the data are retrieved from the JVO portal (http://jvo.nao.ac.jp/portal) operated by the NAOJ. This work was facilitated at the Aspen Center for Physics, which is supported by National Science Foundation grant PHY-1607611.
  \end{acknowledgements}

\bibliographystyle{aasjournal}
%\bibliography{/Users/nienke/Dropbox/Research/myrefs.bib}

\begin{figure}[!ht]
\centering
\includegraphics[width=\textwidth]{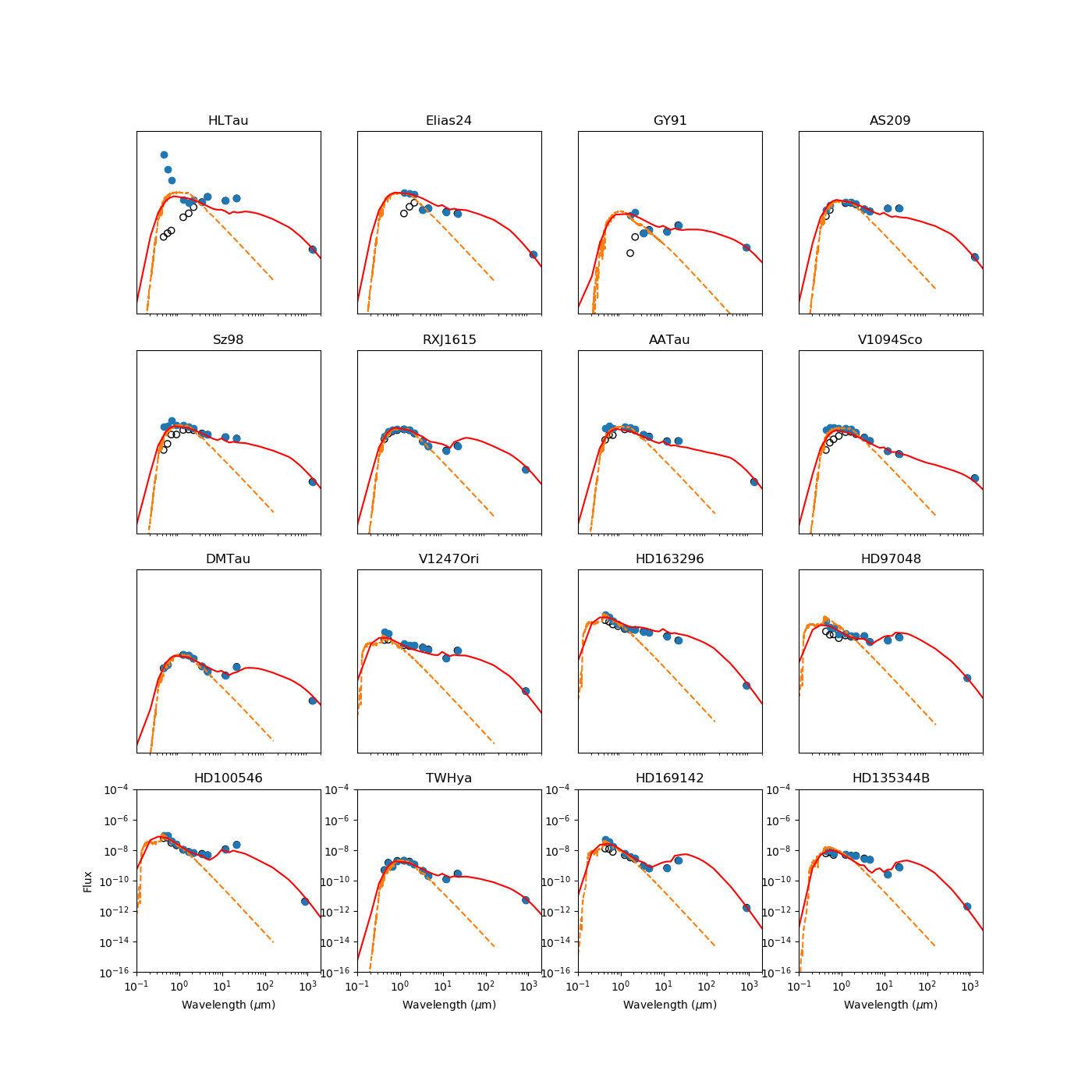}
\caption{Spectral Energy Distributions of each target with their corresponding radiative transfer models. }
\label{fig:SEDs}
\end{figure}

\appendix
Figure \ref{fig:SEDs} presents the Spectral Energy Distributions of each target and their corresponding best-fit model to the dust surface density, that has been used to derive the temperatures in the disk. 

\end{document}